\documentclass[aps,prl,reprint,superscriptaddress,nofootinbib,floatfix,longbibliography]{revtex4-1}
\usepackage[dvipsnames]{xcolor}
\usepackage{amssymb,amsmath,amsfonts,bm,slashed,soul,ragged2e,graphicx,epstopdf,hyperref,array}
\UseRawInputEncoding
\usepackage{caption,subcaption}
\usepackage{booktabs}
\usepackage{multirow}
\usepackage{makecell}
\usepackage{adjustbox}
\usepackage{pifont}
\usepackage{siunitx}
\usepackage{textcase}
\usepackage{enumitem}
\usepackage{url}
\enlargethispage{\baselineskip}
\hypersetup{colorlinks,linkcolor={blue},citecolor={blue},urlcolor={blue}}
\DeclareCaptionJustification{justified}{\justifying}
\everymath{\displaystyle}

\begin{document}
\title{On the effectiveness of the early introduction of modern physics in school curriculum: the case of the structure of atom versus wave-particle duality}

\author{Somya Swarnkar}
\email{somyaswarnkar@gmail.com}
\affiliation{Kendriya Vidyalaya, Katihar, Bihar 854105, India}

\author{Rittick Roy}
\email{rittickrr@gmail.com}
\affiliation{University of Amsterdam, Science Park 904, 1098 XH Amsterdam, The Netherlands}

\author{Tejinder Kaur}
\email{tkaur868@gmail.com}
\affiliation{The University of Western Australia, 35 Stirling Highway, Crawley, WA 6009, Australia}

\author{David Blair}
\email{david.blair@uwa.edu.au}
\affiliation{The University of Western Australia, 35 Stirling Highway, Crawley, WA 6009, Australia}

\date{\today}

\begin{abstract}
\noindent The dual nature of matter and radiation and the concept of the structure of an atom share a number of key conceptual elements from quantum mechanics. Despite the similarities, we find that the concept of the structure of an atom is well understood by students, in contrast to the wave-particle duality. The study analyzes students' comprehension of these two concepts by conducting a semi-structured focus group interview and questionnaire. Through students' performance in the questionnaire and their descriptive responses, we find that the difficulties in their learning and understandings reflect the treatment of the respective topic in the curriculum. The introduction of the structure of an atom is early and repeated, whereas the dual nature of matter and radiation is introduced late and abruptly. Based on our findings, we propose reforms in the present curriculum that are necessary for an improved way of introducing the concept of modern physics, like wave particle duality, to Indian students.
\end{abstract}
\maketitle

\section{I. Introduction} 
The previous century witnessed a major scientific revolution in physics with two ``\textit{avant-garde}" theories that would form the cornerstone of modern physics: the theory of quantum mechanics and the general theory of relativity. Since their inception, these theories have led to a tectonic shift in the status quo of science and technology with the development of nuclear power plants, particle colliders, quantum computers, global positioning system and major theoretical development such as the understanding of the nature of fundamental particles, the theory of black holes and the history of the universe. 

Despite such drastic changes in the physical understanding of the universe, the curriculum in school systems is primarily based on the classical concepts of physics that dates back to the 19th century and shies away from introducing modern physics concepts in the curriculum. The reason being stated often is that concepts such as quantum physics and relativity cannot be visualised or experienced directly \cite{Velentzas2013, Krijtenburg-Lewerissa2017}. Recent researches and studies in the field of physics education, however, advocates to introduce these concepts at an early stage in school education. Introduction of these concepts at an earlier stage makes it easier for students to make a transition to these concepts in their higher studies, than how it is being done now. But going against the established order comes with criticisms and concerns. A major concern raised in this aspect is that these concepts might be too difficult and non trivial to be understood by school students. However, a number of studies over the past ten years has shown that this is not the case. Two major projects in this respect, Einstein First and ReleQuant, have shown that if introduced properly, with a carefully designed teaching-learning environment, students are capable of understanding these concepts at a much younger age \cite{pitts,kaur1,kersting,bungum,kersting2,zahn}.

India is a pioneer in science and research and with the latest developments of LIGO India, India is moving into the ambitious new era of observational age of physics. Moreover, India has the second highest STEM involvement among students after China \cite{stem}. Hence, it is necessary to reflect upon the fact of whether the students are being provided with a time and age appropriate science curriculum in their school education in India and if so, why does a major percentage of the population lacks the scientific attitude, despite having a huge scientific involvement. More specifically, when it comes to the modern understanding of the physical world, why a majority of the population are either unaware of the advancements or hesitates putting their faith into it. 

The claims for the earlier introduction of modern physics concepts in the early stages of the school curriculum by the aforementioned studies motivated the study to dig deeper into the current understanding of these concepts among the science students and to lay the foundation of the reforms needed in the present physics curriculum. This has been achieved through a comparative study of students' comprehension around two major concepts in the realm of modern physics: the structure of atom (SoA) and the dual nature of matter and radiation (DoMR). The study tries to analyze the key areas of difficulty. With the help of a questionnaire and semi-structured focus group interviews, it has been shown that there exists a disparity in the understanding and difficulties of the two subjects. A syllabus review is also presented to provide context for their learning and learning difficulties in the school curriculum. The study shows how an early and gradual introduction of SoA helps in the normalization of the concept and the construction of knowledge around it, in contrast to the DoMR, which is dealt pretty late and abruptly into the curriculum. The attitudes of the students about learning these concepts have also been reflected.

The following paper would be structured as follows: in Section II we discuss the background of the study in detail. In Section III we provide a brief description of the methodology used to conduct the study, the development of the questionnaire and the focus group interviews. Section III also summarizes the results of the study followed by a discussion of the results in Section IV. We conclude the work with final remarks from the study in Section V.

\section{II. Background of the study and research questions}

The idea of introducing the concepts of modern physics to school students is not new. The Einstein First (EF) project has shown over the past 10 years that young students are not only capable of understanding the concepts of modern physics \cite{kaur1}, but benefits from them reflecting it through an increased impact on STEM involvement among school students \cite{Adams2021}. Similarly, the project ReleQuant, has been working on developing online resources for school students to understand the concepts of general relativity and quantum physics in an interactive and interesting way \cite{bungum,kersting2}. 

A second concern that is raised in the context of the early introduction of modern physics, besides students' comprehensibility of modern physics, is why is it necessary to introduce these concepts at an earlier stage. The first and most obvious objective is to introduce students to the best understanding of the physical world. Speaking specifically of the Indian perspective, we quote the National Curriculum Framework 2005 \cite{ncertncf05} for the objective of a ``good science education"\vspace{0.2cm}

\textit{"Good science education is true to the child, true to life and true to science."}\vspace{0.2cm}

Research has shown that to engage a learner comprehensively physics education needs to be contextual with relevant activities and example that relates the learning to their real lives \cite{Stokking2000, Angell2004, Murphy2006, Bøe2013}. Moreover, studies in this respect have shown that the early introduction of these concepts also has a significant scientific, philosophical, and cultural impact on young minds \cite{Monk1997, Abd-El-Khalick2013}. Quoting A.P.French from UNESCO Teaching School Physics \cite{Henriksen2014}\vspace{0.2cm}

\noindent\textit{``..In trying to be efficient we may destroy the whole feeling for what doing physics is like. And what a student will chiefly remember in later years is the attitude with which it is presented, not the particular facts.."}\vspace{0.2cm}

The socio-cultural constructivism of Vygotsky \cite{10.2307/1421493} teaches us how the learners construct their knowledge through an interaction with the environment around them. Language plays an important role in learning. It is the medium through which the learner communicates to it's environment and makes sense out of it. The language of physics in itself is highly multimodel \cite{Lemke1998} and researches have shown the impact of "talking physics" \cite{Henriksen2010}. This is also reflective in the learning of science. The normalization of earth being round is an example of how an inconspicuous idea can become the status quo and accepted in the ``common" understanding. The child constructs it's concept of the earth around it from the very starting and makes it a part of their common sense. As is rightly said by Einstein\vspace{0.2cm}

\noindent\textit{``Common sense is the collection of prejudices acquired by age eighteen" $\sim$ Albert Einstein}\vspace{0.2cm}

We have chosen these two topics, SoA and DoMR, as from a historical standpoint, these two concepts proved non-trivial for the physicists of the time and required extensive study, both theoretically and experimentally, to arrive at the eventual conclusion. It took humankind a century to go from the very first proposition of atoms as the fundamental unit of all material by John Dalton to the first reliable model of the atom with the Geiger-Marsden experiment. It took 20 more years with the development of quantum mechanics to propose a mathematically and experimentally consistent model of an atom. Similarly, the theory of light has gone through its own share of historical oscillations, from the wave motion in ether in the 19th century to the discrepancies in the theory of electromagnetic radiation with the photoelectric effect in the 20th century. Similarly, physicists would scoff at the idea of the quantum nature of particles for decades to come, including prominent fathers of quantum mechanics like Einstein and Schrodinger. Even after reaching a consistent model of quantum mechanics, the non-uniform distribution of particles in an atom, the concentration of mass leaving 99.99$\%$ empty space, the quantized nature of orbital angular momentum and energy levels, and the existence of dual nature for matter and radiation seemed quite counter-intuitive to the physicists of the time.

Despite having such a similar developmental and historical background, are these topics treated on a similar footing in the present school curriculum? And more importantly, do students have a similar level of understanding of these concepts of modern physics? The study tries to investigate these questions in detail.

We try to highlight the fact among young science undergraduates that how the SoA, despite being a non-trivial concept with elements of quantum mechanics sewed into its understanding, introduced relatively early in the school science curriculum, is normalized well among students. Meanwhile the DoMR, introduced later, remains shrouded in the mystical nature of ``quantum weirdness" and hence, not well understood among students. The study tries to gain a deeper insight into the conceptual understanding of the students by conducting a questionnaire and a semi-structured focus group interview and to delve deeper into their learning experiences and difficulties around these concepts and answer the following research questions:
\begin{itemize}
    \item RQ1: Is there any characteristic differences among the understanding of the students around the two concept: SoA and DoMR. If so how do they reflect themselves in students' understanding and comprehension? 
    \item RQ2: Can the differences be understood in context to the science curriculum and how these subjects have been dealt with throughout their learning?
    \item RQ3: From the findings of RQ1 and RQ2, what development and reforms could be proposed to enrich the learning of the concepts of modern physics?
\end{itemize}

\section{III. Research Methodology}
\subsection{A. Focus group interview and questionnaire}
The study was performed by analyzing the conceptual understanding of students around two concepts of modern physics, SoA and DoMR, with the hypothesis that the early introduction of SoA helps students with a better and easier understanding in the later stages of their education than the DoMR. We have chosen a concurrent research design for a combined qualitative and quantitative approach \cite{10.4135/9781483384436} for the analysis of the understanding of the students. A questionnaire was developed with basic conceptual questions to reflect the absolute knowledge and understanding of the referred concepts among the students, along with semi-structured focus group interviews with the same set of students to dig a little deeper into their insights and experiences of learning these concepts throughout different stages of their education. The questionnaire and the FG interview guide can be found in the appendix. We performed thematic analysis to analyze the results from these interviews. The quantitative results from the questionnaire build up a rather objective and statistical foundation for the study \cite{10.4135/9781483384436}. And on the other hand, the qualitative results from the interview provide a socio-cultural perspective to the study by taking into account the personal experiences of the students and the environment, which has potentially affected the process of learning \cite{Taylor2014}. 

We developed the questionnaire to test the understanding of the most basic concepts in the referred topics. In the case of the structure of an atom, the questions asked the participants if they had an elementary idea of the subatomic particles, the structure of an atom, and the energy emission and absorption taking place within an atom. For wave-particle duality, we kept the questions reflective of students' knowledge about the different experimental results that led to the development of the dual nature hypothesis and the concept of DoMR itself. The questionnaire was designed so that any student who has gone through the senior secondary science curriculum in India should be able to answer it. The questionnaire was divided into three sections: the first section was an acknowledgment to the participant about the study and asked for their consent to be a part of this study, while the second and third sections had questions from SoA and DoMR, respectively. 

The interview was semi-structured, with open-ended questions asking students to elaborate on their understanding of the concepts and also share their personal opinions and views on the topics. The interview was also structured to ask students about their personal difficulties that they might have faced while learning about these concepts. Along with this, the students were asked for their views on the historical and philosophical context of these concepts.

A google form for the volunteered participation along with a consent letter, was circulated. 31 students filled the form offering volunteered participation among which 22 students attended the Focus Group (FG) interviews and 20 students filled the questionnaire. Students were first asked to fill up the questionnaire and then attend the interview. The interview was held online and was recorded for the purposes of data analysis with the consent of the volunteers. 6 interviews were taken with a total of 22 students. The recordings were translated and then transcribed by the primary author themselves.


\subsection{B. Sample Space}
The sample space of students for this study was initial year undergraduate students from different universities across India. This demographic was chosen since the concept of DoMR is introduced to school students towards the end of 12th  standard. Hence, to asses the current school and undergraduate curriculum on DoMR, the best set of students are early undergraduate students (first and second year undergraduates). The students' participation was on voluntary basis. The gender ratio is not very inclusive because of the less volunteered participation from students other than males, with a gender ratio of 9:2 in favor of male students.

\section{IV. Students' response and analysis}

The questionnaire consisted of 10 multiple choice questions and 2 subjective response questions. The responses were graded manually by the co-author of this paper ensuring that responses with appropriate answers are rewarded. For example, the students were given a point for the question, ``Select the most appropriate description of the structure of atom" if they either selected the Bohr's model or the Schrodinger's model. Similarly the subjective questions were also graded. The mean score of the students from the questionnaire in the section of structure of atom was found to be $5.16/6.00$. The mean score for the section of dual nature of matter and radiation was found to be $3.70/6.00$, $\sim28\%$ lower than that seen in case of structure of atom. The mean scores of each section and the distribution of responses among them is summarized in Fig. \ref{fig:comp}.
 

After the commencements of the interviews, thematic analysis on the data was performed manually by the first author of the study from transcribed texts, following the six steps mentioned in \cite{10.1191/1478088706qp063oa}. Summary of the thematic coding can be found in Table \ref{tab:themesC}, table III and table IV.

Data from the responses of students was coded around the key concepts in the students' responses. The codes in Table II represent the sub-concepts that reflected themselves in their responses, and then the codes were centralized around the major themes. Table IV is a concurrent summary of both the questionnaire and the FG interview in terms of the frequency distribution of these themes among the students' responses.

Responses for the structure of atom spanned a number of models and concepts, starting from the classical analogy of the solar system to the quantum mechanical distribution of electron clouds around the nucleus. Students were also asked for their views around the absorption and emission of energy by an atom and the empty nature an atom. Following response is an example of how students reported their understanding and the coding of the same response in terms of, "revolving electrons", "centralized mass in atoms" and "probability distributions".\vspace{0.2cm}

 \noindent\textit{FG-student A: ``Atom is mainly empty, three subatomic particle. Electrons revolve around nucleus in energy shells. Most of the mass of atom lies in nucleus."}\vspace{0.2cm}
 
 \noindent\textit{FG-student B: ``There are no such shells, but there is a region of high probability. There is no predefined shell that exist, it's just the high probability region that we call as shells."}\vspace{0.2cm}
 
The responses in the case of DoMR were a little less structured and were difficult to code to their central themes. The major ideas reported spanned from the experimental results of interference/photoelectric effect to the superposition of quantum mechanical states for the case of electron. Misconception, as used by Oslen (2002, \cite{Oslen2002}) is an umbrella term coined for the responses that were either incorrect statements or were major misconception around the concept of dual nature. These misconception were also reflected in the questionnaire as presented in Fig. \ref{fig:pie}\textcolor{blue}{(a)} and Fig. \ref{fig:pie}\textcolor{blue}{(b)}. An example:\vspace{0.2cm}

\noindent\textit{FG-researcher: ``Does matter also have wave nature associated with them?"}\vspace{0.2cm}

\noindent\textit{FG-student: ``Yes mam. For example us, when we do exercise our body gets heated and our body emits infrared waves or heat waves and that's how our body is related with waves."}\vspace{0.2cm}

Along with the conceptual questions and responses among the students, the focus group interviews were also structured to ask for their attitudinal responses and to reflect upon their experiences of learning these concepts. These responses were coded around the central factors that either fostered their interest in physics and positively shaped their learning experience or elements that were reported to be challenging or had an adverse effect on their learning. Another theme that came up often was their experience with learning and taking exams for the same. The attitudinal responses' thematic coding  can be seen in Table \ref{tab:themA}.\vspace{0.2cm}

\noindent\textit{FG-researcher: ``How did you feel learning about these concepts?"}\vspace{0.2cm}

\noindent\textit{FG-student A: ``Mam, plum pudding model was easy but Bohr's model was a little difficult."}\vspace{0.2cm}

\noindent\textit{FG-student B: ``Mam things started contradicting each other. For example a friend asked me if light is a solid or liquid or gas. And one friend asked me how does the transfer of the electrons happens?"}\vspace{0.2cm}

\noindent\textit{FG-student C: ``Our known concepts were challenged all of a sudden. Electrons were particles, and light was a wave. Suddenly, both are both."}\vspace{0.2cm}

\begin{table*}[!t]
\def\arraystretch{1.5}
\setlength{\tabcolsep}{10pt}
\begin{tabular}{c|c|c}
\hline\hline
\textbf{Sections} & \textbf{Themes} & \textbf{Codes}\\
\cline{1-3}
\multirow{5}{*}{Structure of atom} & Rutherford's model & \makecell{$\circ$ Revolving electrons\\$\circ$ Centrifugal forces}\\
\cline{2-3}
& Bohr's model & \makecell{$\circ$ Orbital Picture\\$\circ$ Quantization of momentum}\\
\cline{2-3}
& Schrodinger's model & \makecell{$\circ$ Electron Clouds\\$\circ$ Probability distributions\\$\circ$ Standing waves of electrons}\\
\cline{2-3}
& Atom is mostly empty & \makecell{$\circ$ Repulsive forces\\$\circ$ Concentrated mass at the centre}\\
\cline{2-3}
& \makecell{Quantization of energy and angular momentum} & \makecell{$\circ$ Quantization of energy\\$\circ$ Electron movement from one shell to another\\$\circ$ Energy absorption in discrete value}\\
\cline{1-3}
\multirow{5}{*}{Wave-particle duality} 
& \makecell{Experimental results}  & \makecell{$\circ$ Interference and wave nature\\$\circ$ Photoelectric effect and particle nature\\$\circ$ Other Experimental evidences}\\
\cline{2-3}
& Misconceptions & \makecell{$\circ$ Incorrect statements\\$\circ$ Don't remember\\$\circ$ Empty responses}\\
\cline{2-3}
& \makecell{Quantum mechanical picture} & \makecell{$\circ$ Probability waves\\$\circ$ Superposition of states\\$\circ$ Particle nature and exertion of force}\\
\cline{2-3}
& \makecell{Philosophical and
personal opinions} & \makecell{$\circ$ Neither wave nor particle\\$\circ$ Need for a better theory}\\
\hline\hline
\end{tabular}
\caption{Summary of the thematic coding for the conceptual understanding of the students: The thematic analysis was done following Braun, V. and Clarke, V six steps \cite{10.1191/1478088706qp063oa}. The accounts of the students were coded around the central themes of SoA and DoMR.}
\label{tab:themesC}
\end{table*}

\begin{table*}[!t]
\def\arraystretch{1.5}
\setlength{\tabcolsep}{10pt}
\begin{adjustbox}{width=1\textwidth}
\begin{tabular}{c|c|c}
\hline\hline
\textbf{Sections} & \textbf{Themes} & \textbf{Codes}\\
\cline{1-3}
\multirow{4}{*}{Attitudinal themes} 
& Difficulties and negative influences & \makecell{$\circ$ Contradicting concepts\\$\circ$ Lack of experiments/hands on activities\\$\circ$ Not intuitive\\$\circ$ Mathematical difficulties}\\
\cline{2-3}
& Helpful stuffs positive influences & \makecell{$\circ$ Historical context \\$\circ$ Experimental exercises and activities\\$\circ$ Internet resources}\\
\cline{2-3}
& Exams & \makecell{$\circ$ Practicing questions\\$\circ$ Exam oriented teaching-learning\\$\circ$ Minimal contribution to conceptual understanding}\\
\hline\hline
\end{tabular}
\end{adjustbox}
\caption{Summary of the themes emerging out of the attitudinal responses: Students were asked a number of question to reflect upon their experience of learning SoA and DoMR. Their responses have been coded around the negative or positive influences each theme had on their learning.}
\label{tab:themA}
\end{table*}

\begin{table*}[!t]
\def\arraystretch{1.5}
\setlength{\tabcolsep}{10pt}
\begin{adjustbox}{width=1\textwidth}
\begin{tabular}{c|c|c|c|c|c|c|c|c|c}
\hline\hline
\multirow{2}{*}{\textbf{Students}} & \multicolumn{5}{c|}{\textbf{Structure of atom}} & %
    \multicolumn{4}{c}{\textbf{Wave-particle duality}}\\
\cline{2-10}
 & \makecell{Rutherford's\\model} & \makecell{Bohr's\\model} & \makecell{Schrodinger's\\model} & \makecell{Atom is\\ mostly empty} & \makecell{Quantization\\of energy} & \makecell{Experimental\\results} & \makecell{Probability\\Wave } & \makecell{Misconceptions} & \makecell{Philosophical and\\
personal opinions}\\
\cline{1-10}
 Student 1 & \ding{51} & & &\ding{51} &\ding{51} &\ding{51} & & \ding{51} & \\ 
 Student 2 & &\ding{51} & &\ding{51} &\ding{51} &\ding{56} & &\ding{51} & \\ 
 Student 3 & & &\ding{51} &\ding{51} &\ding{51} & & &\ding{51} & \\ 
 Student 4 & &\ding{51} & &\ding{51} & & & &\ding{51} & \\
 Student 5 & &\ding{51} & &\ding{51} &\ding{51} &\ding{51} & &\ding{51} & \\
 Student 6 &\ding{51} & & &\ding{51} &\ding{51} & & &\ding{51} &  \\
 Student 7 & &\ding{51} & &\ding{51} &\ding{51} &\ding{51} & &\ding{51} & \\
 Student 8 & & &\ding{51} &\ding{51} &\ding{51} &\ding{51} & &\ding{51} &  \\ 
 Student 9 & & &\ding{51} &\ding{51} &\ding{51} & &\ding{51} & & \\ 
 Student 10 &\ding{51} & & &\ding{56} & & & &\ding{51} & \\ 
 Student 11 & & &\ding{51} &\ding{51} &\ding{51} & & & \ding{51} & \ding{51} \\ 
 Student 12 &\ding{51} & & &\ding{51} &\ding{51} &\ding{56} & &\ding{51} & \\ 
 Student 13 & & &\ding{51} &\ding{51} &\ding{51} & &\ding{51} & & \\ 
 Student 14 &\ding{51} & & &\ding{51} &\ding{51} &\ding{51} & &\ding{51} & \\ 
 Student 15 & &\ding{51} & &\ding{51} &\ding{51} & &\ding{51} & & \\ 
 Student 16 &\ding{51} & & &\ding{51} & \ding{51}& & &\ding{51} & \\ 
 Student 17 &\ding{51} & & &\ding{51} &\ding{51} &\ding{56} & &\ding{51} & \\ 
 Student 18 & &\ding{51} & & \ding{51} &\ding{51} & \ding{51} & & & \\ 
 Student 19 & & &\ding{51} & \ding{51} &\ding{51} &\ding{51} &\ding{51} & &\ding{51} \\ 
 Student 20 & & &\ding{51} & \ding{51} &\ding{51} &\ding{51} &\ding{51} & &\ding{51} \\ 
 Student 21 & &\ding{51} & &\ding{51} &\ding{51} &\ding{56} & \ding{51} & & \\
 Student 22 & & &\ding{51} &\ding{51} &\ding{51} &\ding{56} & \ding{51} & & \\
\hline\hline
\end{tabular}
\end{adjustbox}
\caption{Student wise analysis of the themes: The table is a combined analysis of the student's accounts from the questionnaire and the focus group interview. The students were consistent with their responses. Whenever a student has reported any account falling into any above mentioned themes or if their questionnaire responses adhered to that theme a \ding{51} has been marked, whenever a student reported the response and it was an incorrect response a \ding{56} has been marked. The empty spaces denote a case of no response from the student.}
\label{tab:studdata}
\end{table*}

\begin{table}[!t]
\def\arraystretch{1.5}
\setlength{\tabcolsep}{10pt}
\begin{tabular}{c|c}
\hline\hline
Section & Mean score\\
\cline{1-2}
Structure of atom & 5.16 (86\%)\\
Wave-particle duality & 3.70 (61\%)\\
\hline\hline
\end{tabular}
\caption{The mean score (out of a total score of 6) of all the students from the questionnaire across the two sections}
\label{tab:score}
\end{table}

\subsection{A. Analysis of students' response}

\begin{figure}
\centering
\includegraphics[width=0.45\textwidth]{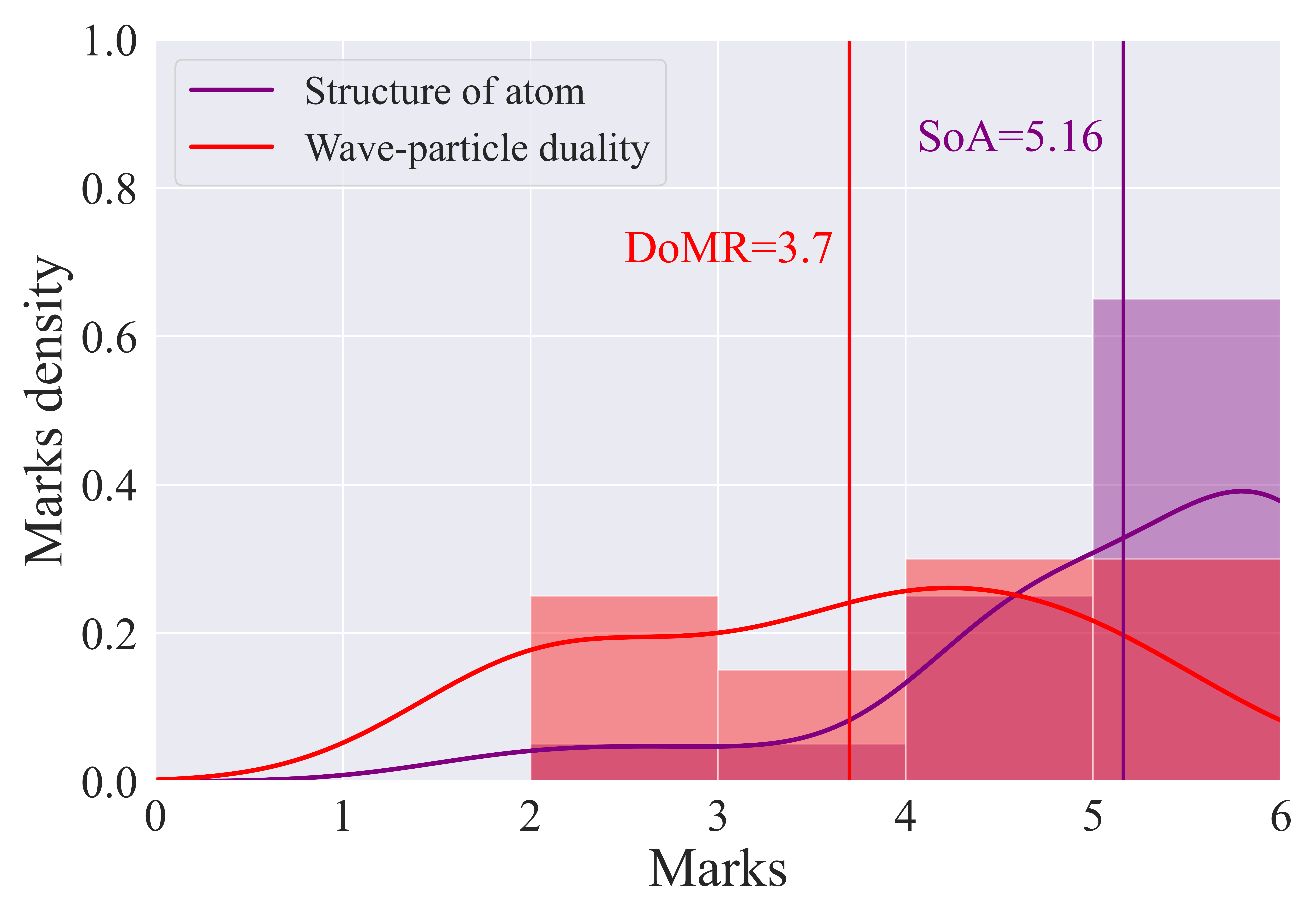}
\caption{The plot shows the kernel density estimate and the associated histograms of the marks density for the distributed questionnaire. The red (purple) plot and the histogram shows the marks distribution for the students for the section of DoMR (SoA). The mean score is shown with the vertical lines for each section.}
\label{fig:comp}
\end{figure}

\begin{figure*}
		\centering
		\subfloat[What does the peak and trough of an electron wave denotes? ]{{\includegraphics[width=0.21\textwidth]{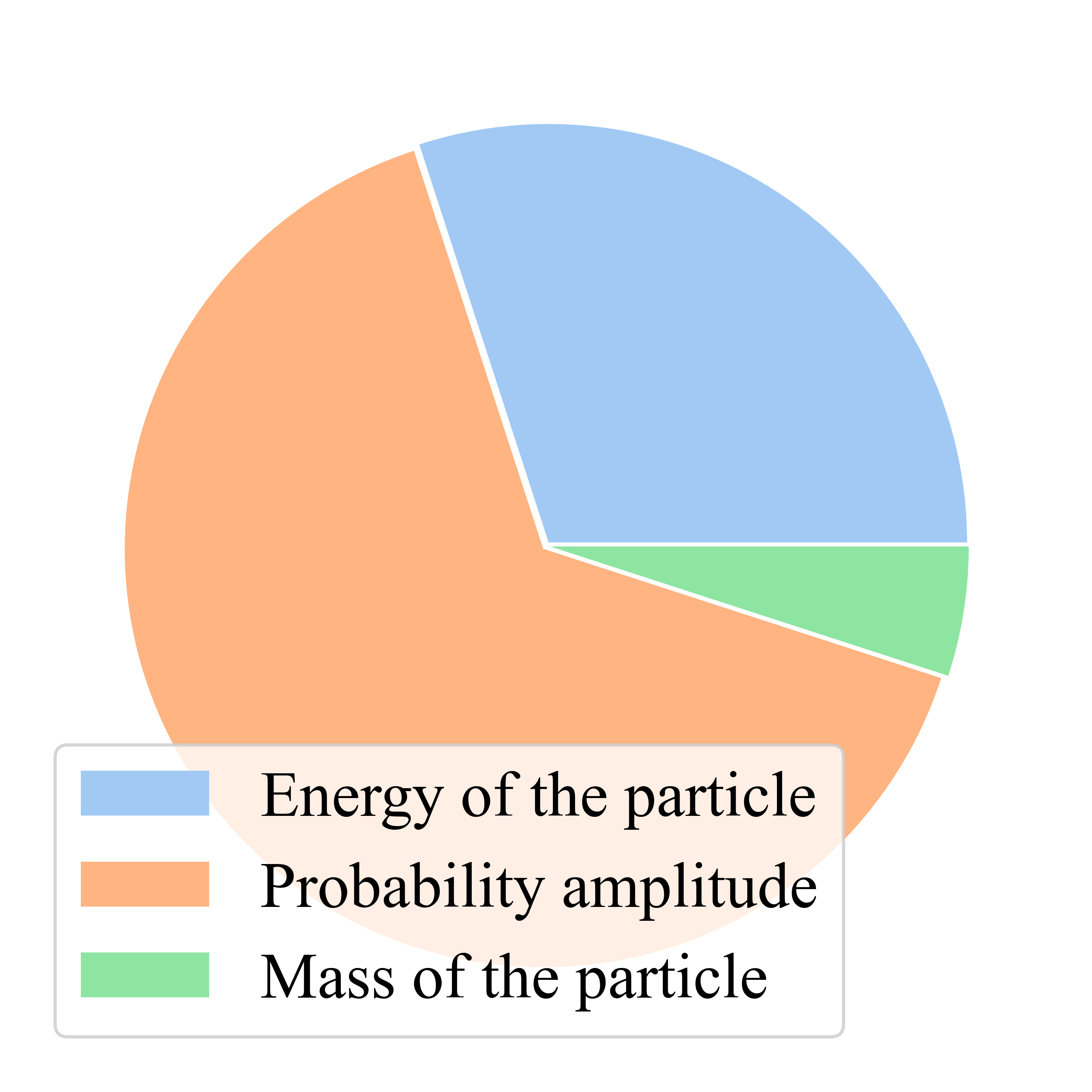} }}%
		\qquad
		\subfloat[What does the \\wave nature of a \\particle corresponds to?]{{\includegraphics[width=0.21\textwidth]{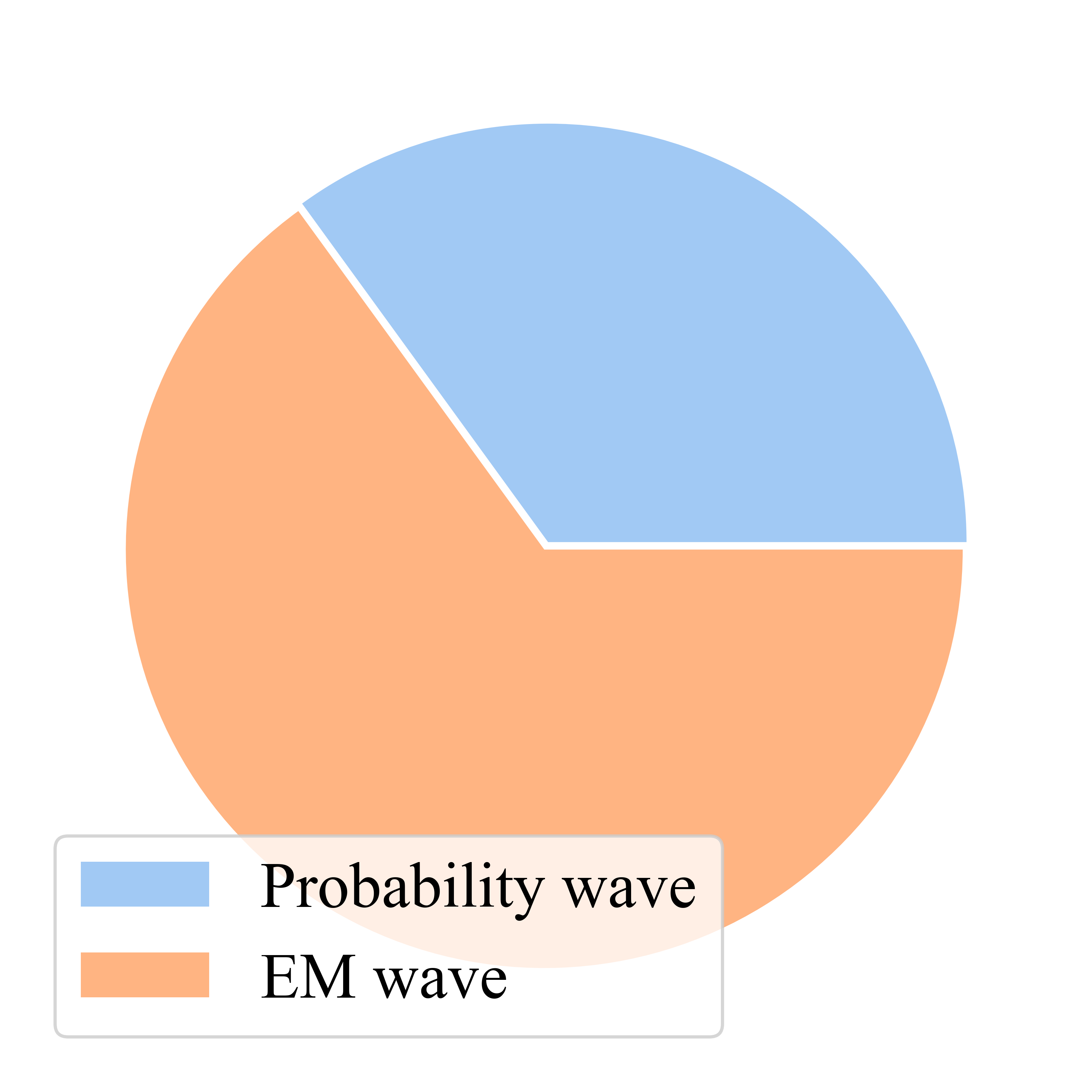} }}%
		\subfloat[How does an atom emit or absorb radiation?]{{\includegraphics[width=0.21\textwidth]{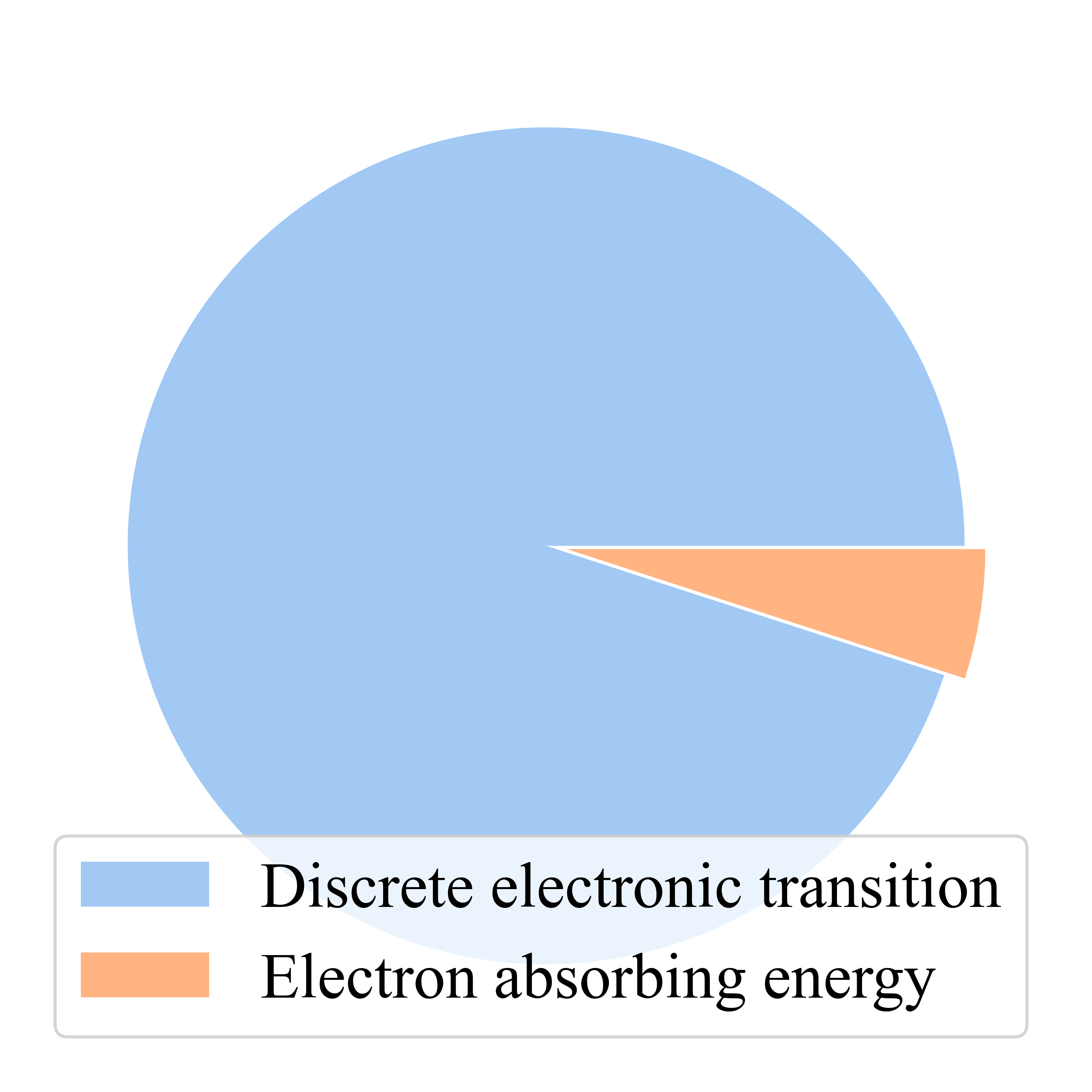} }}%
		\qquad
		\subfloat[Select the most appropriate description of the structure of atom.]{{\includegraphics[width=0.21\textwidth]{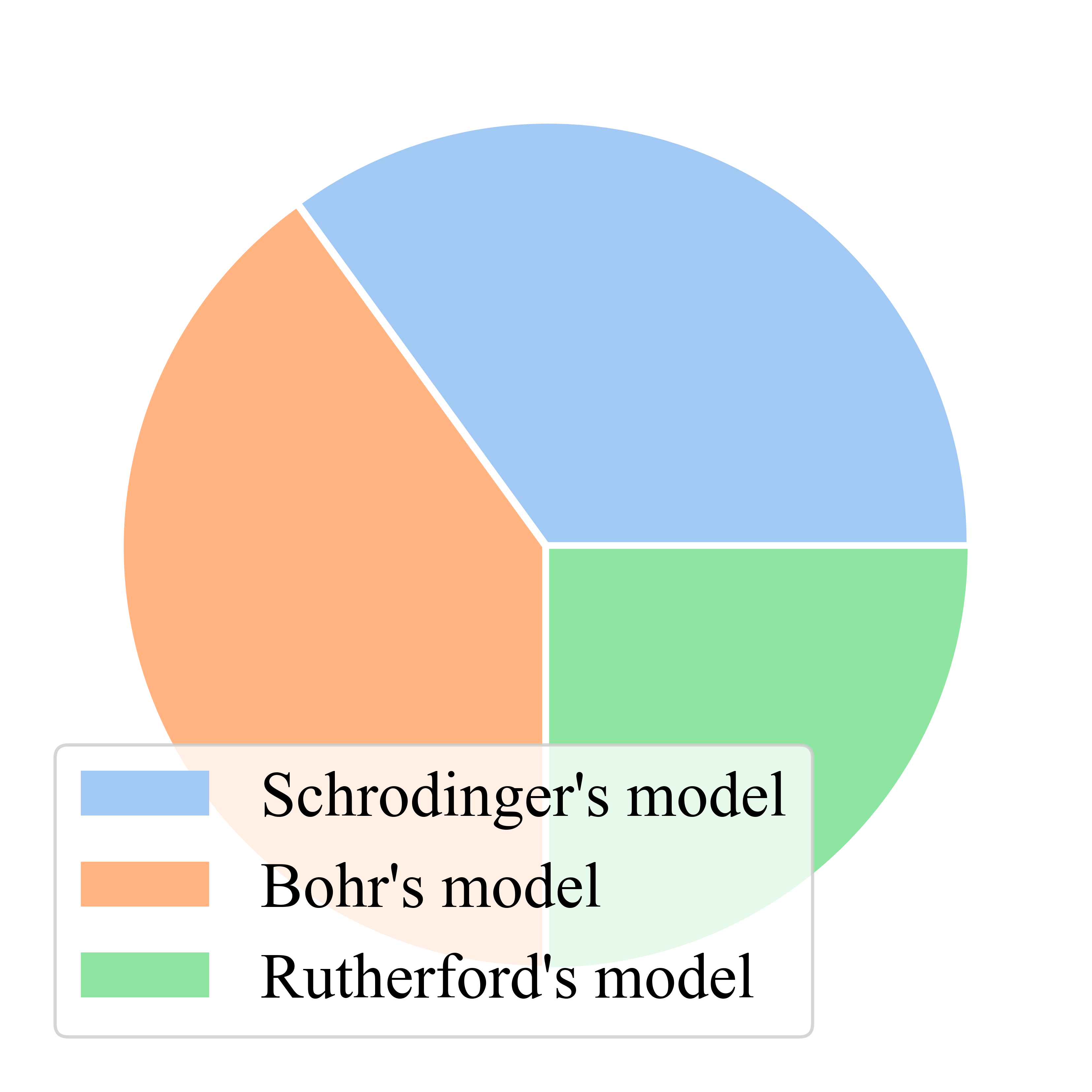} }}%
		\caption{Pie-chart showing the distribution of answers across four selected questions from the questionnaire.}%
		\label{fig:pie}%
\end{figure*}

This study was aimed towards investigating the students' understandings and difficulties around two concepts in modern physics, and reflect upon the differences in the learning because of the different ways they have been introduced and dealt throughout the school and undergraduate curriculum. The results from the questionnaire and the thematic analysis makes this difference evident. 

The difference in the mean scores from the questionnaire (3.70 for DoMR and 5.16 for SoA, refer Table \ref{tab:score}) reflects on the contrast in  students' retention of the respective concept and their ability to reproduce the same. Fig. \ref{fig:comp} summarizes the distribution of the scores of the student and shows that the associated kernel density estimate (KDE) in the case of SoA is skewed towards the higher end, while that of DoMR is rather flat, representing a consistently better performance in the case of SoA. Table \ref{tab:studdata} shows the distribution of individual student and their respective responses around the central themes. The table is a concurrent summary of the results from both the questionnaire and the focus group interview. The responses for the case of SoA is distributed among different atomic models and almost every student reported a comprehensive understanding for the concentration of mass in the nucleus, quantization of the energy levels and the discrete nature of emission and absorption of energy by electrons. Following are some excerpts from the students which showcases the technical language compatibility among students around SoA:\vspace{0.2cm}

\noindent\textit{FG-researcher: ``Why doesn't the electron fall into the nucleus?"}\vspace{0.2cm}

\noindent\textit{FG-student: ``The orbitals they are in have constant energy, it doesn't gain or lose energy. Its not like a particle revolving, its a cloud of probability, revolution picture is not correct. It is only when we observe its here or there, otherwise it has an equal spread, it doesn't revolves in the modern picture."}\vspace{0.2cm}

It was seen throughout the FG interviews that even when asked about the structure of the atom, students would often mention quantum properties such as the probabilistic distribution of electrons, the quantization of electrons' angular momentum obtained by equating the wavelength to the circumference of the orbit, and the discrete nature in which energy is absorbed or emitted by electrons in the form of photons. Thus, inadvertently, students were seen to be discussing the concept of DoMR while talking about SoA.

On the other hand, the distribution of responses for the case of DoMR, in Table \ref{tab:studdata}, can be seen to be centralized around the experimental observation and misinterpretations. Only 7 students talk about the wave particle duality from a quantum mechanical standpoint, and only 3 of them have enough understanding to go further ahead to present their personal views into the topic apart from the textbook material.\vspace{0.2cm}

\noindent\textit{FG-student: ``Wave particle duality is taught by giving experiment and saying it cannot be explained by one so (it) must be the other. So it's just the way we are trying to explain (wave particle duality), both should be valid."}\vspace{0.2cm}

It was a common observation throughout the FG interviews that students were inclined to mention the experimental evidence for the wave-particle duality of electromagnetic radiation as soon as they were asked about the DoMR. Students didn't always have the retention as to which experiment highlighted the wave or particle nature, but their responses reflected their understanding of DoMR to be built around the experimental results. Students exhibited language compatibility around the experimental results, but when asked about further insights, their skewed conceptual understanding of the wave nature and the particle nature was evident. When asked ``What do we mean when we say light has a wave nature?", almost 30$\%$ of the students reported that ``light needs a medium to travel", along with a mention of the interference of light. Similarly, when asked to reflect on their understanding or visualization of the wave nature of particles, most of the students either had empty responses or major misconceptions about the topic. Fig. \ref{fig:pie}\textcolor{blue}{(b)} shows how the majority of the students responded to electromagnetic waves when they were asked about what the wave nature of particles corresponds to. The results comply with the studies \cite{Oslen2002, Mannila2001} that suggest students' misconceptions are rooted in a classical physics worldview.\vspace{0.2cm}

\noindent\textit{FG-student A: ``I liked SoA. DoMR was not understood with clarity. We still don't understand how to see the waves or understand waves. We know how to calculate values but I don't really understand how does it happens."}\vspace{0.2cm}

\noindent\textit{FG-student B: ``Light travels in a straight line, which also tells us about it's wave nature."}\vspace{0.2cm}

An important difference to note here is that while discussing SoA, students were seen to have a comprehensive understanding of the particle nature of waves and the wave nature of particles, as discussed earlier. Fig. \ref{fig:pie}\textcolor{blue}{(c)} shows the percentage of students who mentioned the discrete nature of electromagnetic radiation, thus talking about the particle nature of waves. At the same time, when discussing these concepts in the realm of DoMR, students were seen struggling with the theory. Fig. \ref{fig:pie}\textcolor{blue}{(a)} and Fig. \ref{fig:pie}\textcolor{blue}{(b)} show the flawed understanding of DoMR among students. But more interestingly, when asked ''What do the peak and trough of an electron wave denote?", about 65$\%$ of students answered that it is the probability amplitude, but when asked ``What does the wave nature of a particle correspond to?", only 35$\%$ students answered probability wave, showing that there is a lack of consistency in their answers, reflecting an incomplete understanding of DoMR. 

\subsection{B. Attitudinal Responses}
While collecting their attitudinal responses, students were asked to choose one topic that they felt was easy to understand. Of the 22 students, 13 chose SoA, 4 chose DoMR, 1 student said none as they found both on equal footing, and 4 students didn't respond. The thematic analysis of the attitudinal responses reflects the factors that might have had a positive or negative influence on the learner. One major factor that came up too often was how the historical and contextual information helps the students not only relate to the concept but also construct the concepts in their minds around its needs, discoveries, experiments, and conclusions. \cite{olivia, Arriassecq2012}\vspace{0.2cm}

\noindent \textit{FG-student: ``There was a curiosity to understand SoA. Also the history that was told about the structure of atom helped a lot in understanding them. It started very basic."}\vspace{0.2cm}

The majority of the students mentioned that they needed explicit help, either in the form of other books, internet searches, or video explanations, to understand DoMR when having difficulties. In the entire sample space, 4 students reported still being curious about these concepts and either pursuing their career forward in this avenue or learning about these concepts independently.\vspace{0.2cm}

\noindent\textit{FG-researcher: ``Do you think learning about these concepts affected your interest in physics?"}\vspace{0.2cm}

\noindent\textit{FG-student A: ``I was more reliable on Youtube and it fostered my curiosity. I still am curious about these things and like to learn them about more each and every day."}\vspace{0.2cm}

\noindent\textit{FG-researcher: ``Tell us more about your experience in quantum mechanics?"}\vspace{0.2cm}

\noindent\textit{FG-student B: ``My first introduction was through Youtube videos. We learnt about this in high school. Rest of that I used to follow YouTube. We also had a course in college about quantum Mechanics."}\vspace{0.2cm}

Another aspect that was reported heavily in the sessions was the learner's experiences with different examinations. Students who have gone through the process of preparing for competitive exams talked about the exposure to the rigor of solving plenty of mathematical problems. This contributed to their confidence in mathematics and physics. But at the same time, the fact that this didn't contribute to their conceptual understanding was also reported by some students. On the other hand, some students reported the difficulties they face because of the lackluster way modern physics is being introduced in the curriculum, leaving them with an array of open questions, leading to poor performance in their exams and a decreased interest in physics.\vspace{0.2cm}

\noindent \textit{FG-student: ``It was very difficult in the initial stages, after solving many problems I got used to it. Dual nature wasn't intuitive, so started solving problems, not understood theoretically."
}

\section{V. Syllabus review}
In this section we present a review of the syllabus pertaining to the topics in consideration, i.e. SoA and DoMR, throughout the school and undergraduate curriculum in India. In this study, we would follow the Central Board of Secondary Education (CBSE) for the syllabus of school level physics and the syllabus set by the University Grants Commission (UGC) for the undergraduate level physics in India. 

In order to analyse the key-features of the school curriculum of SoA and DoMR in the school science curriculum, we analysed the textbooks of standard 9-12, published by the National Council of Educational Research and Training (NCERT). NCERT is an autonomous organisation set up in 1961 by the Government of India, and is responsible for publishing textbooks and study material for the school students in India. Most of the schools and Education Boards in India refer to NCERT Books. 
In the school curriculum, the concept of an atom is first introduced in standard 9 \cite{ncerttextbook}. 
\vspace{0.2cm}

\textit{``What is an Atom?
Have you ever observed a mason building
walls, from these walls a room and then a
collection of rooms to form a building? What
is the building block of the huge building?
What about the building block of an ant-hill?
It is a small grain of sand. Similarly, the
building blocks of all matter are atoms.."}\vspace{0.2cm}

Introducing atoms as the building blocks of matter around us, various models of the atom in their respective historical context is discussed along with the shortcomings of every model, with a subtle emphasis on the experimental validity of the models presented. The chapter concludes with a description of Bohr's model of an atom.\vspace{0.2cm}

\textit{``In order to overcome the objections raised
against Rutherford’s model of the atom,
Neils Bohr put forward the following
postulates about the model of an atom:
(i) Only certain special orbits known as
discrete orbits of electrons, are allowed
inside the atom.
(ii) While revolving in discrete orbits the
electrons do not radiate energy"}\vspace{0.2cm}

Following this introduction to SoA, a brief discussion is presented for the discovery of neutron and the orbital structure of atom. The SoA is reintroduced to students in 11th standard, but this time with a quantum mechanical description of electrons around the nucleus. 

The concept of waves is first introduced in 10th standard in the chapter of ``Sound". Properties of waves such as frequency, wavelength and amplitude is discussed in detail. We note that this introduction comprises only of pressure waves, since the context in which this is introduced is that of sound waves.\vspace{0.2cm}

\textit{``A wave is a disturbance that moves through a medium when the particles of the medium set neighbouring particles into motion. They in turn produce similar motion in others. The particles of the medium do not move forward themselves, but the disturbance is carried forward. This is what happens during propagation of sound in a medium, hence sound can be visualised as a wave.."}\vspace{0.2cm}

Waves are discussed later in 11th standard physics text as an explicit chapter where the mathematics of wave mechanics gets introduced. Electromagnetic radiation and the propagation of oscillating and magnetic field gets introduced in 12th standard, before the discussion of wave optics. 

The concept wave particle duality gets introduced in 12th standard. After a detailed discussion of the wave nature of light along with fundamental wave-like properties such as interference and diffraction, experimental evidences are presented that requires a particulate description of light such as the photoelectric effect. This marks the introduction of DoMR. Following this discussion, the dual nature of matter is discussed, albeit with less detail. The material concludes with a discussion about the Davission-Germer experiment introducing the wavelike nature of electrons.

For the undergraduate syllabus, we analyzed two books on introductory quantum mechanics \cite{Beiser, eisberg}, suggested as textbooks for the undergraduate curriculum by UGC. The coursework begins with an introduction of the failures of classical theory by presenting some key experimental results showcasing the particle nature of electromagnetic radiation: Planck's law on black-body radiation and Einstein's theory of the photoelectric effect. This is followed by a discussion on the Compton scattering of electrons, introducing the wave nature of particles. With the experimental results presented, the uncertainty principle and the DoMR is introduced. A brief discussion is presented regarding the consequences of the DoMR. These postulates are then used to motivate the development of the Schrodinger's equation. Students are introduced to different techniques to solve the Schrodinger's equation for simple quantum mechanical systems. After the introduction of the basic postulates of quantum mechanics, the different models for the SoA are introduced and discussed in detail. The curriculum concludes with the calculation of the energy eigenstates and the electron wavefunction of simple atoms such as that of the Hydrogen atom and discussing the implications of the existence of discrete energy eingenstates.  

The syllabus summary is presented in Table \ref{tab:syllabus}. The table includes the topics covered with respect to their time of introduction in the curriculum, subject, and context. Till 10th grade, students do not have the segregation of science as physics, chemistry, or biology. It happens in the 11th standard, only for those who chose the science stream.

SoA is dealt with in a structured and gradual fashion, enriched with the historical context of its development. From its first introduction in the 9th standard to the quantum mechanical discussion in the 12th standard, it follows a structured and repetitive pattern. On the other hand, the introduction of DoMR is rather brief and abrupt. Also, the context of the very first introduction of the wave nature of particles follows a very classical approach to sound waves. After that, the wave theory gets its place in a scattered and nonstructured way. The first quantum mechanical description of DoMR in the 12th standard seems abrupt and all at once. There's a profound difference in the curriculum about both concepts in terms of their first introduction, context, historical progression, and experimental proofs.


\begin{table*}[!t]
\def\arraystretch{2}
\setlength{\tabcolsep}{15pt}
\begin{adjustbox}{width=1\textwidth}
\begin{tabular}{c|c|c|c}
\hline\hline
\textbf{Curriculum} & \textbf{Standard} & \textbf{Structure of atom} & \textbf{Wave-particle duality}\\
\cline{1-4}
\multirow{4}{*}{School curriculum} & Standard 9 & \makecell{$\circ$ Thomson's model\\$\circ$ Rutherford's model\\$\circ$ Bohr's model\\$\circ$ Electrons in shells (K,L,M,N..)\\$\circ$ Valency} & ---\\
\cline{2-4}
& Standard 10 & \makecell{---} &\makecell{---}\\
\cline{2-4}
& Standard 11 & \makecell{(Chemistry)\\$\circ$ Schrodinger's model\\$\circ$ Pauli's exclusion principle\\$\circ$ Quantization of momentum\\$\circ$ Different energy levels\\and electron's
movement in that} & \makecell{(Chemistry)\\$\circ$ Dual nature of matter\\and radiation discussed very\\briefly in the chapter for SoA}\\
\cline{2-4}
& Standard 12 & \makecell{(Physics)\\$\circ$ Electron Clouds\\$\circ$ Probability distributions\\$\circ$ Standing waves of electrons\\$\circ$ Quantization of energy} & \makecell{(Physics)\\$\circ$ Wave nature of Light\\$\circ$ Interference and diffraction\\$\circ$ Photoelectric effect\\/particle nature of light\\$\circ$ Davission and Germer experiment\\/wave nature of electrons\\$\circ$ Interference of light and matter\\and the
understanding of\\wave nature associated\\$\circ$ Photoelectric effect and\\particle nature of light
and electrons\\$\circ$ Experimental evidences reflecting\\the need of
dual nature}\\
\cline{1-4}
\multirow{1}{*}{Undergraduate curriculum} 
& \makecell{1st year} & \makecell{$\circ$ Review of atomic models\\$\circ$ Bohr's model and postulates\\$\circ$ Sommerfield's model\\$\circ$ Solution of one electron systems\\$\circ$ Energy eingenstates and\\wavefunction interpretation of H-atom}  & \makecell{$\circ$ Discrepancy with black body spectrum\\and photoelectric effect\\$\circ$ Planck's theory of black body\\$\circ$ Einstein's theory of photoelectric effect\\$\circ$ Compton effect\\$\circ$  Matter waves\\$\circ$ Dual nature of matter and radiation}\\
\hline\hline
\end{tabular}
\end{adjustbox}
\caption{The list of subject matter and topics, and when they have been introduced in the curriculum. For school syllabus we have referred to the syllabus prescribed by the CBSE with NCERT textbooks. The undergraduate syllabus has been taken from the UGC prescribed syllabus for the undergraduate physics course.}
\label{tab:syllabus}
\end{table*}




\section{VI. Discussion}
The results from the FG interviews and questionnaires can be very well understood if we put them in perspective with the curriculum around these concepts. The wave-particle duality was introduced in the school syllabus in the 12th standard. The learning begins with a dedicated chapter on interference and diffraction, which discusses the wave nature of light in detail. This is followed by a discussion about the photoelectric effect in detail, discussing the particle nature of light. Experimental results are mentioned in detail, but there is very little or no elaboration on the facts about what it really means to have a wave or particle nature and how they contradict each other. In the earlier standards, the concept of wave is introduced in the context of sound waves and how they need a medium to travel. The drastic transition to electromagnetic waves and their wave/particle duality are not dealt with carefully, leaving a void for the student to make their own sense of these concepts, resulting in misconceptions or leaving unanswered curiosities, resulting in either empty or no responses. 

On the other hand, the SoA has remained a recurring concept in the curriculum since the 9th grade. The introduction in the 9th standard gives them comprehensive knowledge about the atomic model, building the concepts of centralized mass and electrons being in certain ``shells" or ``orbits". The subject is re-introduced in the 11th standard, with a more detailed description of its quantum mechanical character. But despite the recurring nature of the concept, the fact that 14 out of 22 students responded with either the classical Rutherford's model or Bohr's model supports the constructivist hypothesis in the early years of a student's career.

\noindent \textit{FG-student: ``We keep discarding the different models of SoA but the Bohr's model wasn't discarded at that moment which made us believe Bohr's atomic model is the ultimate truth and it sticks with us for longer times. We all have studied about atomic clouds and probability distribution but still many of us imagine the Bohr's model for SoA because it retained very much"}\vspace{0.2cm}

Students didn't report the SoA being very difficult, as the concepts were not contradictory but rather constructing around the structure of atoms. On the other hand, putting the wave particle duality later in the curriculum takes away the opportunity for construction and lays down the foundation for cognitive conflicts and contradictions around the concerned concept.

Based on the above results, we propose the following improvements and ways to deal with the subject matter in a more comprehensible fashion:
\begin{itemize}
    \item The early introduction of the concepts of modern physics will help in the construction of knowledge in an accurate and more comprehensible way. Throughout the study, the socio-cultural constructivism theory of learning has presented itself in many shapes and forms. The early years of a child's education contribute to the construction of their worldview, which will manifest throughout their lives. Hence, this should be the time for the introduction of the most appropriate understanding of the physical world. They should be taught to appreciate and take benefit of the historical developments that led to the destination, but they need not take the same journey.
    \item Concepts should be introduced with their most accurate description, simplified, and tailored to their age-appropriate cognitive development. EF project has repeatedly shown this in several studies over the past 10 years \cite{kaur1}. The curriculum and specific lesson plans could be curated to make the content cognitively competent for the learner. The focus should be not to overburden them but, also not to introduce them to older or less accurate scientific content.  
    \item Hands-on activities and appropriate models should be used in the visualization of these concepts. Online resources are already helping students. Conscious and careful development of such online resources will significantly improve teaching and learning. As students have reported taking help from resources other than the textbook and course material, in this new age of technology we should take advantage of the vast range of resources that can be developed to aid the visualization of the rather abstract concepts \cite{Müller2002,Lee2010,Ainsworth2006,Plass2009,Kluge2010}. ReleQuant has developed an array of such open online resources for several concepts in modern physics. Likewise, these multimedia resources must be combined with research-based teaching and learning strategies \cite{Singh2008}.
\end{itemize}
\section{VII. Conclusion}
We live in an era where the exponential growth of science has been so normalized that nothing seems impossible anymore. Children don't see the awe in stepping on the moon, and knowing the location of your loved ones with the vast network of satellites has become a necessity. And yet, modern-day understanding in the commons about the universe around them lacks the depth and accuracy that are needed for the next generation to take science forward and appreciate the nuances and beauty of the world that we live in.

The study aimed to answer the questions that revolved around the understandings and conceptions of two modern physics concepts among science students by assessing the motivations and challenges from the perspective of the school and undergraduate physics curricula. A contrast between two important modern-physics concepts, the structure of the atom and the dual nature of matter and radiation, was proposed because of the early and recurring introduction of the one while brief and abrupt discussion of the other.

With a combination of a questionnaire and semi-structured focus group interviews, the study glides into students' understanding and reflects the disparity between the two: SoA and DoMR. The relation between their alternative conceptions and the treatment of the concepts in the curriculum, emerges from the thematic analysis conducted on students' accounts from the interview. 

The curriculum of physics in India, for senior secondary students, does talk about quantum mechanical concepts such as the Schrodinger model of the structure of an atom, the dual nature of matter and radiation, and Heisenberg's uncertainty principle. However, a contrast study of the two topics reflects the disparity between the treatment in the secondary school curriculum. On one hand, the SoA was introduced in the 9th standard with a qualitative description of the historical developments of different models, ending its note by presenting Bohr's model. The chapter takes the reader on a historical journey by discussing major scientific milestones, such as the discovery of subatomic particles, in the context of the development of SoA. The later section presents a vague idea of electron shells and the filling of electron shells with the rule of $2n2$. Students' accounts from the interviews and their responses to the questionnaire show that the curriculum reflects itself in their understanding. Although students were introduced to the modern-day Schrodinger model in their senior secondary curriculum, most students tend to retain Bohr's model or the classical Rutherford's model, as it was the last model taught to them in their secondary school years.

DoMR has not been treated with appropriate care and detail. Students were seen to report classical interpretations of the wave nature of radiation. The students did talk about the experimental evidence of the photoelectric effect and interference to mention the origin of the problem, but when asked further open-ended questions about the wave nature of particles, misconceptions in their understanding were pretty evident.

The reason for this could be attributed to the fact that the curriculum might consider the topic cognitively challenging for the students, hence not giving it the needed attention. But at the same time, SoA, also a quantum mechanical and non-trivial concept, gets its needed attention, and students can be seen showing their compatibility around the concept.

Students are able to reproduce textbook accounts of DoMR, but when asked further open-ended questions, they can be seen to struggle and report major misconceptions. The students had difficulty understanding the wave and particle nature of matter and radiation, which reflected the lack of connections around the concept. On the other hand, most students can be seen as more comfortable and responsive around the SoA, reflecting the normalization of the concept in the socio-cultural setting. 

The results support the constructivist hypothesis of Vygotsky that the students construct their knowledge through their interactions with the environment. The concepts, taught with needed care and attention in the early years of their schooling, get constructed in their minds as such, and the children make sense of them naturally. Subject matter taught in the early years of schooling makes their world of common sense difficult to challenge later in their lives. Most of the students are still reporting Bohr's model, although all of them have been introduced to the quantum mechanical model, which is evident of this very fact.

This study proves that if students are capable of understanding the SoA, they are also capable of understanding modern-day physics concepts provided the needed attention to the subject matter in the curriculum. We anticipate more studies to follow in the Indian context to develop interactive online and offline curricula to teach modern physics concepts and to see the applicability of the same in classrooms. We believe that students deserve the best understanding of the contemporary world, and we need to work towards this.

\section{ACKNOWLEDGMENTS}

The authors express their gratitude to Magdalena Kersting for valuable discussions in the early phases of the research. The authors would like to thank the participating students in the focus group interviews and questionnaire. All participants voluntarily involved with this study gave their informed consent for the study and publication of the results.  

\section{Appendix A: Focus Group Interview Guide}

\begin{itemize}
    \item \textbf{Conceptual Understanding: } 
	What do you understand by the wave particle duality? What does the wave nature of light correspond to? 
	What do you think experimentally prove the wave nature of light and how?
	How do you visualise an electron having a wave nature?
	
	Can you explain the structure of an atom to me?
	Why doesn't the electron fall in the nucleus? 
	How does an atom absorb energy? What happens when it absorb energy?
	How is an atom of a calcium different from that of an iron?
	Does every material around us have atoms as their building blocks?
	Do you think the living organisms are also made up of atoms?
	
	\item \textbf{Insights into their understandings: } 
	How will you summarize the difference between classical and quantum mechanics?
	Was learning about the structure of atom, easy relative to the dual nature of light and matter?

	\item \textbf{Evaluation of the learning content: }
	The knowledge that you feel to possess about Quantum mechanics, do you think your textbooks and the classroom teaching has major contribution to it? Do you remember learning about it?
	When do you think your were formally exposed the early and basic foundations of it?
	When do you thing the structure of atom was taught to you?
	You remember it being too difficult to understand at that stage?
	When do you think you were taught about the dual nature of matter and radiation?
	Do you remember it being difficult to understand at that stage?

	\item \textbf{Historical and philosophical approach: } 
	Do you really believe all that is said about the dual nature of radiation and matter or say other modern physics concept, that is actually the reality? 
	Do you think learning about these concepts somehow affected your interest in physics?
	Would you appreciate if physics were taught with more conceptual and intuitive insights, rather than having a rigor mathematical and abstract construct?
	Do you think the understanding of the nature of light or how electrons are situated in an atom, is significant to you? 
	
End Note.

\end{itemize}

\section{Appendix B: Questionnaire}

We are trying to analyze the difficulties and challenges faced by the students in their
introduction to Modern physics concepts. Here we have put forward some conceptual
questions to know more about your understanding about these concepts. 

\subsection{A. Acknowledgement}

This questionnaire is developed to analyze your understanding about two major Modern physics concepts: Dual Nature of matter and radiation and Atomic physics. This is a part of a study being conducted to analyze the difficulties and challenges faced by students in their introduction to Modern Physics in Schools. Your answers and responses would be used for research purposes only. \textbf{Your responses would be treated confidentially, you would not be identified by name or recognized otherwise in the research reports. } Your name and email address are needed to identify different responses separately and to maintain the reliability of the study. The questionnaire is graded for the data analysis purposes only.

\begin{enumerate}
    \item Please enter your name.
    \item Please enter your email address
    \item I have read the above acknowledgement carefully and I understand the purpose of this questionnaire explicitly. I give my consent to use the responses provided by me for the research purposes only.
    \begin{itemize}
        \item I agree
    \end{itemize}
    \item I understand that my responses might effect the result of research and hence I will be truthful about my answers and responses with the best of my knowledge.
    \begin{itemize}
        \item I agree
    \end{itemize}
\end{enumerate}

\subsection{B. Structure of Atom}

\begin{enumerate}
    \item Select the subatomic particles (Multiple Select Question)
    \begin{itemize}
        \item Photons
        \item Electrons
        \item Protons
        \item Neutrons
    \end{itemize}
    \item Smallest constituent element of everything around us are (Choose the most appropriate answer)
    \begin{itemize}
        \item Atoms
        \item Molecules (which are a different category of constituent particles other than atoms) and atoms
        \item Molecule (which is a group of two or more atoms that are chemically bonded together) and atoms.
        \item Atoms and molecules for non living things, and cells for living organisms.
    \end{itemize}
    \item What makes an atom of iron different from that of a calcium?
    \begin{itemize}
        \item One is stronger than the other
        \item Iron has the same atom bigger in size, while that of the calcium is smaller
        \item They have different number of electrons protons and neutrons
        \item Iron and Calcium has same atom. It's the electron distribution in different orbitals which makes them different.
    \end{itemize}
    \item Atom’s are mostly empty, and most of the mass of an atom is concentrated in it's nucleus.
    \begin{itemize}
        \item True
        \item False
    \end{itemize}
    \item Can you explain a little as to how we can understand the thermal properties of a material by considering the atomic picture? (How do you understand temperature and pressure with the help of atomic picture?)
    \item How does an atom emit or absorb radiation?
\end{enumerate}

\subsection{C. Dual Nature of Particles}

\begin{enumerate}
    \item How does very small quantities such as electrons and photons behave as?
    \begin{itemize}
        \item Waves
        \item Particles
        \item Either waves or particles
        \item Both waves and particles
    \end{itemize}
    \item What does we mean when we say light has a wave nature? (You can select more than one option)
    \begin{itemize}
        \item It exhibits wave like characteristics such as interference in experiments.
        \item It can transfer energy and momentum from one place to another.
        \item Light needs a medium to travel
        \item It spreads everywhere in the space.
    \end{itemize}
    \item What does the wave nature of particle corresponds to?
    \begin{itemize}
        \item EM wave
        \item Probability wave
        \item Gravitational wave
        \item Pressure wave
    \end{itemize}
    \item Does Earth has a wave associated with it just like photons or electrons?
    \begin{itemize}
        \item Yes
        \item No
    \end{itemize}
    \item What does the peak and trough of an electron wave denotes?
    \begin{itemize}
        \item Energy of the particle
        \item Trajectory of the particle
        \item Mass of the particle
        \item Probability amplitude of the particle
    \end{itemize}
    \item Suggest an experimental result that can be used to show the wave | particle nature of photons.
    \begin{itemize}
        \item Young’s double slit | photoelectric effect
        \item Diffraction |Young’s double slit
        \item Compton Scattering | Photoelectric effect
        \item Compton Scattering | Diffraction
    \end{itemize}
\end{enumerate}

\bibliography{ref.bib}

\end{document}